\documentclass[%
 reprint,
 amsmath,amssymb,
 aps,
]{revtex4-2}

\usepackage{graphicx}
\usepackage{xcolor}
\usepackage{dcolumn}
\usepackage{bm}

\begin{document}

\preprint{APS/123-QED}

\title{A Bi-CMOS electronic-photonic integrated circuit quantum light detector}

\author{Joel~F.~Tasker}
 \affiliation{Quantum Engineering Technology Labs, H. H. Wills Physics Laboratory
and Department of Electrical \& Electronic Engineering, University of Bristol, BS8 1FD, United Kingdom}
\author{Jonathan~Frazer}%
\affiliation{Quantum Engineering Technology Labs, H. H. Wills Physics Laboratory
and Department of Electrical \& Electronic Engineering, University of Bristol, BS8 1FD, United Kingdom}
\author{Giacomo~Ferranti}%
\affiliation{Quantum Engineering Technology Labs, H. H. Wills Physics Laboratory
and Department of Electrical \& Electronic Engineering, University of Bristol, BS8 1FD, United Kingdom}
\author{Jonathan~C.~F.~Matthews}
\email{Jonathan.Matthews@bristol.ac.uk}
\affiliation{Quantum Engineering Technology Labs, H. H. Wills Physics Laboratory
and Department of Electrical \& Electronic Engineering, University of Bristol, BS8 1FD, United Kingdom}

\date{\today}

\begin{abstract}
Complimentary metal-oxide-semiconductor (CMOS) compatible quantum technology enables scalable integration with the classical readout and control electronics needed to build quantum computers. Homodyne detectors have applications across quantum technologies including quantum computers, 
and they comprise photonics and electronics. Here we report a quantum noise limited monolithic electronic-photonic integrated homodyne detector, with an overall footprint of $80~\mu\mathrm{m} \times 220~\mu\mathrm{m}$, fabricated in a 250~nm lithography bi-polar CMOS process. By monolithic integration of the electronics and photonics, overall capacitance is suppressed---this is the main bottleneck to high bandwidth measurement of quantum light. We measure a 3~dB bandwidth of 19.8~GHz and a maximum shot noise clearance of 15~dB. This exceeds bandwidth limits of detectors with macroscopic electronic interconnects, including wirebonding and flip-chip bonding. This demonstrates CMOS electronic-photonic integration enhancing performance of quantum photonics.
\end{abstract}

\maketitle

Photonic integrated circuits (PIC) are a compelling approach to develop quantum technology~\cite{WangNatphoton2020,Moody_2022} and they underpin proposed architectures for optical quantum computing~\cite{RudlophOptimisticAPLP,XanaduBlueprint}. CMOS compatible PIC platforms, such as silicon on insulator photonics~\cite{si-jstqe-22-390,GiewontIEEEJSTQE2019}, offer paths to scaling up the manufacture of photonic devices for quantum technology in commercial foundries. This may prove critical in the construction of universal quantum computers, because the scale and performance required of components to build quantum computers is beyond anything yet constructed in information technologies~\cite{RudlophOptimisticAPLP}.

Since initial experiments with silicon quantum photonic circuits~\cite{si-natphot-2014,GentryOptica2015}, CMOS compatibility for electronic-photonic integration has been a clear goal for quantum photonics. This is because it would enable integration at scale of components generating and utilising quantum states of light with the required high performance classical readout and control electronics. But to date, the development of foundry ePIC platforms~\cite{ihp_epic,GiewontIEEEJSTQE2019} has been driven by the performance demands of classical applications, with demonstrations including 56~GB/s direct detection receivers~\cite{epic1} and 128~Gb/s coherent receivers~\cite{epic2} for fibre optics telecommunciations, and coherent detector arrays with active pixel amplifiers for 3D imaging~\cite{Rogers5123DimagerChip}.
    
\begin{figure}[tp!]
        \centering
        \includegraphics{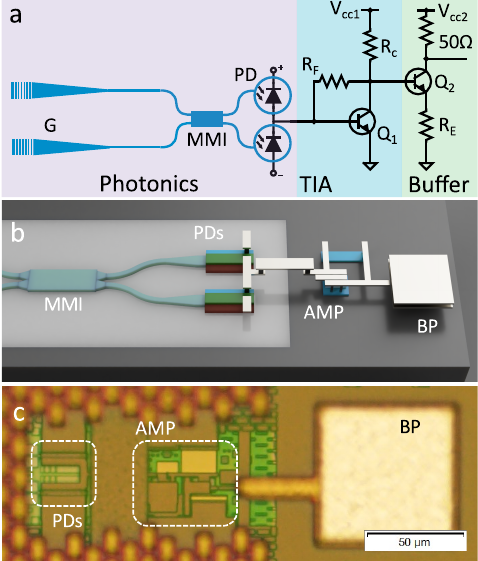}
        \caption{\textbf{A Bi-CMOS integrated homodyne detector for measuring quantum light.} \textbf{a} The detector schematic. The photonics include grating couplers (G), mode converters, strip waveguides, a multi-mode interference coupler beamsplitter (MMI) and germanium-silicon photodiodes (PDs). The electronics are a two-stage TIA design. The first transistor (Q$_1$) forms a common-emitter shunt-feedback TIA; the second, (Q$_2$), constitutes a $50~\Omega$ output buffer amplifier. R\textsubscript{F}, R\textsubscript{C}, R\textsubscript{E} label the feedback, load and emitter resistors. \textbf{b} A 3D illustration of connections between components using three of the five metal layers in the SG25H5 EPIC process~\cite{ihp_epic} used to fabricate the device. Light grey indicates silicon-on-insulator, dark grey indicates bulk silicon. \textbf{c} A microscope image of the detector illustrates scale. AMP labels the TIA and buffer amplifier stages. This device fits within a 80~$\mu$m~$\times$~220~$\mu$m footprint.}
        \label{fig:concept_fig}
    \end{figure}

Here we demonstrate electronic-photonic integration can be applied to enhance quantum technologies. We report integration in one monolithic ePIC chip (Figure~\ref{fig:concept_fig}) of all the electronics and silicon photonics needed for homodyne detection of quantum optical signatures~\cite{lvovsky2001quantum}. The detector has a measured 3-dB bandwidth of $19.8\pm0.1$~GHz and a maximum measured shot noise clearance of 15~dB. By extrapolating the measured clearance, we infer shot noise limited performance beyond the bandwidth of our analysis equipment, measuring more than 10~dB at 26.5~GHz.

Homodyne detectors can measure weak signals by interfering them with a local oscillator at an optical beamsplitter. The resulting interference is observed in the subtraction of photocurrents from a pair of photodiodes placed at the two beamsplitter outputs. This subtraction current requires amplification, and when the amplification electronics are of sufficiently low noise, the homodyne detector is sensitive enough to reveal quantum noise signatures in the input. This is quantified by the clearance between optical shot noise and the electronic noise of the detector. Quantum technology applications of homodyne detectors include squeezed-light-enhanced gravitational wave detection~\cite{PhysRevLett.123.231107,Acernese2019Dec}, quantum state tomography~\cite{lvovsky2001quantum}, measuring continuous variables cluster states~\cite{Larsen369,Asavanant373} for quantum computing and for continuous variables quantum communication~\cite{LodewyckPRA2007CVQKD}. 

Waveguide integrated beamsplitters have been used for homodyne detection with silica-on-silicon ~\cite{ma-natphot-316-9} and lithium niobate PICs~\cite{Lenzinieaat9331}. In silicon-on-insulator photonics, on-chip germanium p-i-n photodiodes have been integrated with waveguides and interfaced with discrete amplifier electronics for quantum random number generation and coherent state tomography~\cite{raffaelli2018homodyne}, and as a chip-scale receiver for continuous variables quantum key distribution~\cite{zh-natphot-13-839}. In these cases, the detector bandwidths were limited to respectively $\sim$100~MHz and $\sim$10~MHz by discrete electronics, mounted on printed circuit boards (PCB). Consequently, micro-electronic amplifiers were wirebonded to silicon PICs, and the resulting detectors demonstrated 3-dB bandwidths of 1.7~GHz~\cite{Tasker2021Jan} and {1.5~GHz}~\cite{Bruynsteen:21} -- these detectors were respectively used to measure squeezing over a 9~GHz bandwidth and observe shot noise clearance out to 20~GHz. 

A remaining limiting factor in the speed of these detectors is the 20 fF - 100 fF capacitance overhead of the electrical bondpad interconnection~\cite{soicmos}, that interfaces the PIC with the integrated electronics. Flip-chip interfaces introduce similar capacitance overheads, and so also restrict the possible bandwidth of hybrid integration using macroscopic interconnects. In order to increase bandwidth further, monolithic integration is required.

The reported single-chip homodyne detector is illustrated in Figure~\ref{fig:concept_fig}. It was designed and characterised in-house with fabrication outsourced to the Leibniz Institute for High Performance Microelectronics (IHP). We chose IHP's SG25H5\_EPIC process, which features a 250~$\mu$m silicon node, germanium-based photodiodes with $f_{3\textrm{dB}} > 60$~GHz and vertically integrated heterojunction bipolar transistors (HBTs) for RF applications using 250~nm lithography with a specified transition frequency $f_T = 220$~GHz and a breakdown voltage of 1.7 V~\cite{ihp_epic}. The RF performance of these HBTs is comparable to the lateral n-channel MOSFET transistors in references~\cite{GFIBM,luxtera,Monolithic15Oracle_Buckwalter_2012}. This is due to the vertical carrier transport of the HBT, meaning speed is less dependent on the lithography resolution allowing vertical bipolar transistors to outperform NMOS devices at the same process node~\cite{RetrospectiveOnSiGe_Cressler_2013}. The HBTs are integrated in the same front-end-of-line process as the silicon-on-insulator waveguides and active optical components, such as modulators and photodiodes. This approach removes all bondpad and packaging parasitics, with connections between photonics and electronics made in the metal interconnect layers of the back-end-of-line (BEOL).

The IHP fabrication process begins with a SOI wafer optimised for photonics, with a 220nm silicon layer thickness and a 2um thick buried oxide layer. A ‘local-SOI’ approach is employed in which SOI regions that are to be used for BiCMOS devices are etched down to the silicon substrate. Bulk silicon is selectively regrown epitaxially in these regions and is subsequently planarised using chemical-mechanical planarisation. Patterning of electronic and photonic structures is conducted in parallel and the electrical contacts to the photodiodes and transistors are made with the same process step~\cite{ihplocalsoi}. Devices are then connected through a single shared BEOL with five metal layers. 

The transimpedance amplifier (TIA) used consists of a HBT common-emitter amplifier in shunt-feedback configuration, followed by a 50~$\Omega$ buffer amplifier for interfacing with standard radio frequency (RF) test equipment (see Figure~\ref{fig:concept_fig}). The bandwidth of a single-stage shunt-feedback TIA with an ideal second-order Butterworth response is given by~\cite{sackinger_limit}
    \begin{equation}
    \label{eq:transimpedance_amp}
       f_{3\mathrm{dB}} =\sqrt{ \frac{A_0 f_A}{2 \pi C_{\mathrm{in}} R_F }},
    \end{equation}
where $C_{\mathrm{in}}$ is the total capacitance at the amplifier input, $R_F$ is the feedback resistance and $A_0 f_A$ is the gain-bandwidth product. A monolithic design reduces $C_{\mathrm{in}}$ by minimising the stray capacitance between the photodiodes and amplifier due to bondpads or other wiring related sources. This comes in addition to the already low capacitance associated with integrated photodiodes and high performance HBTs -- integrated photodiodes with capacitances as low as 9~fF and amplifier input capacitance of order 100~fF have been reported~\cite{VokicTIA_InputC,Lischke:15}. This is in stark contrast to the packaging and layout associated parasitic capacitance on a PCB of up to tens of picofarads~\cite{masalov2017noise,raffaelli2018homodyne}. Eq. \ref{eq:transimpedance_amp} demonstrates the fundamental trade-off between the detector bandwidth and the transimpedance gain from the subtraction photocurrent to output voltage. Larger transimpedance gains are desirable to ensure the detector noise lies above the noise floor of any subsequent equipment and provide the maximum shot noise clearance when a local oscillator field is applied. However, the practically usable $R_F$ and achievable bandwidth are constrained by the total input capacitance and the gain-bandwidth product. In the case of a single transistor amplifier, the gain-bandwidth product is proportional to the transistor transition frequency, $f_T$ via $A_0 f_0 \approx C_I/ C_L f_T$, where $C_I/C_L$ is ratio of transistor input and load capacitances~\cite{Sackinger2017AnalysisReceivers}.

The input-referred current noise power spectral density is given by,
    \begin{eqnarray}
    \label{eq:transimpedance_current noise}
    I_{n, T I A}^2(f) &=& \frac{4 k_B T}{R_F} + \frac{2 q I_C}{\beta} + 2 q I_C \frac{\left(2 \pi C_T\right)^2}{g_m^2} f^ 2\\ && +~4 k T R_b\left(4 \pi C_{P D}\right)^2 f^2,  \nonumber
    \end{eqnarray}
where $I_C$ is the HBT collector current, $\beta$ is the DC current gain, $g_m$ is the transistor transconductance and $R_b$ the base resistance \cite{Sackinger2017AnalysisReceivers}.  The first two terms are white noise terms, specifically the feedback resistor Johnson noise and base current shot noise, respectively. The latter terms scale quadratically with frequency to a limit set by the photodiode junction capacitance and total capacitance, including parasitics, presented to the amplifier input. 

The amplifier is implemented with two n-p-n transistors as shown in Figure \ref{fig:concept_fig} a, the design of which is provided as part of the SG25H5\_EPIC process development kit (PDK). The transition frequency $f_T$ is maximised for a particular collector current density\textemdash for our collector area, this corresponds to an optimal bias current $I_C$ of 4.5~mA. Achieving the optimal collector current requires careful tuning of the biasing resistors $R_C$ and $R_E$ for a given transimpedance gain $R_F$. We perform lumped element SPICE simulations of the amplifier to optimise resistances with $V_{cc1} = 2.2~V$ and $V_{cc2} = 1.7~V$ dictated by the transistor breakdown voltage. The chosen resistors are $R_F = 600~\Omega$, $R_c = 250~\Omega$ and $R_E = 35~\Omega$ where the feedback resistance has been chosen to provide sufficient clearance above the fundamental thermal noise floor of the $50~\Omega$ termination resistor in RF test equipment. Photonic layout was performed using IPKISS and Cadence Virtuoso. Simulations, electronic design, layout and post-layout electronic simulations were performed using Cadence Virtuoso using PDK SPICE models provided by IHP.

The gain spectrum of an ideal shunt-feedback TIA is that of a second-order Butterworth filter, given by
    \begin{equation}
        G(f) = \frac{A^2_0}{1 + \left( \frac{i2\pi f}{i2\pi f_{3\mathrm{dB}}} \right)^2 }
        \label{eq:gain_spectrum}
    \end{equation}
where $A^2_0$ is the absolute gain at zero frequency. The second stage buffer operates as a unit gain amplifier and can be assumed to have a bandwidth approximately equal to the transistor transition frequency \cite{Sackinger2017AnalysisReceivers}.

A 3D model and a microscope image of the detector is shown in Figure \ref{fig:concept_fig} b \& c. A $20~\mu$m trace connects the photodiodes subtraction signal the the amplifier input. Our parasitic extraction simulations estimate the parasitic capacitance of this interface at 7~fF, compared with 105~fF when simulating a single bondpad at the amplifier input. The ePIC is bonded to a purpose-made PCB designed for high-frequency operation. Vertical silicon capacitors (Murata UWSC, 1~nF) are used on the PCB for power supply decoupling on transistor and photodiode biases. The ePIC itself contains additional vertical metal-insulator-metal capacitors located next to each component for additional supply filtering (not shown in Figure \ref{fig:concept_fig} a). The TIA output wirebond is kept short to minimise parasitic inductance.

We characterise the bandwidth, common-mode rejection ratio (CMRR), linearity and responsivity of the device. Light is coupled into the chip using grating couplers, and multimode interferometers (MMIs) are used as beamsplitters.  A continuous-wave (CW) tuneable laser (PurePhotonics PPCL550) at 1550~nm and amplified with an erbium-doped fibre amplifier (PriTel), is used as a local oscillator (LO). A variable optical attenuator (VOA, OzOptics) adjusts LO power. Noise measurements are recorded using a Keysight N9020B MXA electronic spectrum analyser (ESA) with a 26.5~GHz bandwidth. Photodiode and transistor biases are supplied from sourcemeters (Keysight U2722A \& Keithley 2450) which are also used to monitor the two individual photocurrents of the diodes. We compare measured photocurrents when injecting LO at the top and bottom MMI ports, finding a splitting ratio of 42:58 transmission to reflection. This imbalance results in a net photocurrent at the amplifier input and excess electronic noise at the amplifier output (see Appendix). We offset this effect by reducing the bias on the bottom photodiode until the photocurrents are matched, with a maximum difference of $80~\mu$A at the maximum LO power. This reduces the quantum efficiency of the bottom diode to $72\%$ of its maximum value.

The top and bottom photodiodes are each reverse biased at 2~V and -0.3~V, respectively, relative to an amplifier input voltage of 0.9~V. The transistor supplies, $V_{cc1}$ and $V_{cc2}$ (see Figure \ref{fig:concept_fig} a), are set to 2.2~V and 1.65~V. To account for signal loss from PCB transmission lines and coaxial cables, we measure S21 parameters of a PCB co-planar waveguide test structure and the coaxial cable used in the experiment using a Keysight N5225A network analyser.

\begin{figure}
        \centering
        \includegraphics[scale = 0.9]{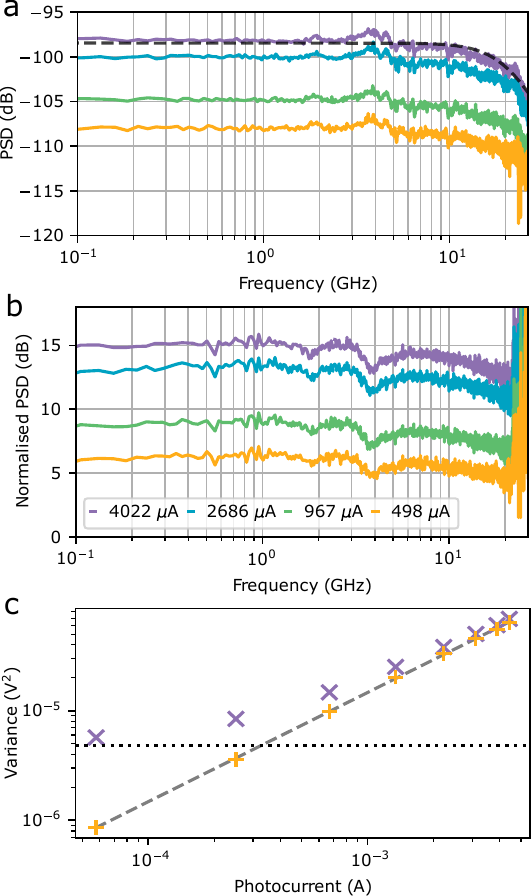}
        \caption{\textbf{Homodyne detector characterisation.} \textbf{a}, Power spectral density (PSD) of the detector where ESA noise and the amplifier dark noise have been subtracted in addition to cable and transmission line loss corrections. The legend represents the total photocurrent measured on both photodiodes. The dashed line shows a fit to Eq~\ref{eq:gain_spectrum} and gives a 3~dB bandwidth of $19.8 \pm 0.1$~GHz. \textbf{b}, PSD of the detector normalised to the amplifier electronic noise. \textbf{c}, Raw and electronic noise subtracted detector noise variance at 1~GHz against total photocurrent. The horizontal line represents the electronic noise level. A linear fit to the data (dashed) indicates a gradient of $0.99\pm0.01$, demonstrating the presence of vacuum shot noise up to a maximum clearance of 15~dB. }
        \label{fig:spectra}
    \end{figure}

We perform a bandwidth measurement by optimising coupling at maximum power using the monitored photocurrent, then recording a series of spectra on the ESA as the VOA adjusts the input power from 13.5~dBm to -26.5~dBm. We also record the ESA displayed average noise level (DANL, the intrinsic ESA noise) for later subtraction from the data. All spectra are recorded at 100~kHz RBW over a 26.5~GHz span. The results of this are plotted in Figure \ref{fig:spectra}. By fitting the detector response to a second-order Butterworth response, we obtain a 3-dB bandwidth of $19.8\pm0.1$~GHz. As the clearance of the detector extends beyond the bandwidth of our ESA, we estimate the shot noise bandwidth using Eq. \ref{eq:transimpedance_current noise}. We fit the clearance of the detector with $A/(B + Cf^2) + 1$ where $A$ describes the optical shot noise, $B$ the white noise terms of Equation \ref{eq:transimpedance_current noise} and $C$ the latter frequency dependent terms (see Appendix). The fit suggests that the shot noise clearance extends far beyond the measured bandwidth, vanishing beyond 100~GHz. In practice we anticipate the photodiode transit time bandwidth to become limiting~\cite{Liu1999Jul}. 

Grating coupler losses are measured using grating-to-grating test structures that we included on the ePIC chip. This yields an average of approximately $4.0~$dB per coupler.
We characterise the photodiode responsivity by comparing the sum of measured photocurrents to off-chip LO power and correcting for grating coupler losses. From this, we obtain a maximum photodiode responsivity of 0.47~A/W at 2~V bias, including MMI insertion loss. 

\begin{figure}[htp]
    \centering
    \includegraphics[scale = 0.88]{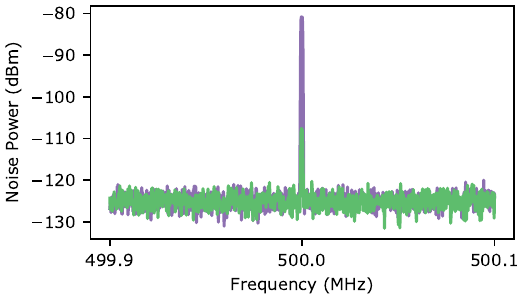}
    \caption{\textbf{Detector common mode rejection ratio at 500~MHz}. The LO power is set to generate $10~\mu$A of total photocurrent and the noise power recorded with one or both photodiodes reverse biased. We observe a maximum of 27~dB CMRR at 500~MHz}
    \label{fig:cmrr}
\end{figure}

CMRR measurements are made by intensity-modulating the LO using an electro-optic modulator and comparing the signal with either both photodiodes biased as above, or one biased and the other disconnected to eliminate its photocurrent contribution. The ESA is set to 10~kHz RBW and a span $\pm 0.1\%$ of the modulation frequency. We observe a CMRR of 27~dB at 500~MHZ (Figure~\ref{fig:cmrr}), which is limited by the intrinsic splitting ratio of our MMI. In future devices, this value can be improved by substituting the current static MMIs with thermoelectric tuneable Mach-Zehnder interferometers~\cite{Tasker2021Jan}. 

An ePIC quantum light detector is reported, combining photonics and readout electronics within a 80~$\mu$m~$\times$~220~$\mu$m footprint. This was achieved thanks to the CMOS compatibility of silicon photonics, which can benefit the scalability and manufacturability of photonic quantum information processors and could be a potential necessity when considering the stringent timing limits imposed by feed-forward and delay lines~\cite{scott2022timing}. The detector's $19.8\pm0.1$~GHz 3-dB bandwidth is an order of magnitude greater than previous fastest demonstrations and surpasses the speed performance limits of homodyne detectors constructed from macroscopic wirebond interconnects~\cite{Tasker2021Jan}. The demonstration maintained shot noise efficiencies of at least 95\%. Higher gains, and thus higher efficiencies, will be possible in future devices through multi-stage amplifier designs without sacrificing bandwidth~\cite{sackinger_limit}. Higher responsivity photodiodes have been demonstrated in silicon photonics, achieving 95\% quantum efficiency with 30~GHz bandwidths in classical appliations~\cite{Benedikovic:19}, and fibre-coupling efficiencies of 95\% have been observed with edge couplers~\cite{BakirIEEE2010}. Incorporating such improvements will enable ePIC detectors to simultaneously meet all of the performance requirements of future quantum technologies. We believe the current detector's footprint and performance already opens application of ePIC homodyne detectors to minaturised and high speed receivers for quantum communications~\cite{LodewyckPRA2007CVQKD,zh-natphot-13-839}, higher clock rates cluster state characterisation~\cite{Larsen369,Asavanant373} and large arrays of coherent receivers for continuous variables photonic quantum computing~\cite{XanaduBlueprint} and photonic neural networks operating below the Landauer limit~\cite{PhysRevX.9.021032}. 

Beyond detectors, we anticipate future applications of ePICs to increase the performance of quantum device control, including increasing the number of simultaneously controlled phase shift parameters beyond O($10^2$) in highly programmable quantum devices~\cite{vi-natphys-17-1137}. We expect the combination of minaturised readout and control within ePICs will reduce the requirements on optical delay lines for quantum technologies utilising state-measurement and feedforward~\cite{scott2022timing}. This is important for large-scale implementations of quantum technology including, multiplexed sources of quantum states~\cite{Xiong2016Mar}, quantum state engineering~\cite{CableDowlingPRLNOON2007} and measurement-based and time-multiplexed quantum computing~\cite{XanaduBlueprint,Takeda2017Sep}.

\bibliography{References}

\providecommand{\noopsort}[1]{}\providecommand{\singleletter}[1]{#1}%
\begin{thebibliography}{44}%
\makeatletter
\providecommand \@ifxundefined [1]{%
 \@ifx{#1\undefined}
}%
\providecommand \@ifnum [1]{%
 \ifnum #1\expandafter \@firstoftwo
 \else \expandafter \@secondoftwo
 \fi
}%
\providecommand \@ifx [1]{%
 \ifx #1\expandafter \@firstoftwo
 \else \expandafter \@secondoftwo
 \fi
}%
\providecommand \natexlab [1]{#1}%
\providecommand \enquote  [1]{``#1''}%
\providecommand \bibnamefont  [1]{#1}%
\providecommand \bibfnamefont [1]{#1}%
\providecommand \citenamefont [1]{#1}%
\providecommand \href@noop [0]{\@secondoftwo}%
\providecommand \href [0]{\begingroup \@sanitize@url \@href}%
\providecommand \@href[1]{\@@startlink{#1}\@@href}%
\providecommand \@@href[1]{\endgroup#1\@@endlink}%
\providecommand \@sanitize@url [0]{\catcode `\\12\catcode `\$12\catcode
  `\&12\catcode `\#12\catcode `\^12\catcode `\_12\catcode `\%12\relax}%
\providecommand \@@startlink[1]{}%
\providecommand \@@endlink[0]{}%
\providecommand \url  [0]{\begingroup\@sanitize@url \@url }%
\providecommand \@url [1]{\endgroup\@href {#1}{\urlprefix }}%
\providecommand \urlprefix  [0]{URL }%
\providecommand \Eprint [0]{\href }%
\providecommand \doibase [0]{https://doi.org/}%
\providecommand \selectlanguage [0]{\@gobble}%
\providecommand \bibinfo  [0]{\@secondoftwo}%
\providecommand \bibfield  [0]{\@secondoftwo}%
\providecommand \translation [1]{[#1]}%
\providecommand \BibitemOpen [0]{}%
\providecommand \bibitemStop [0]{}%
\providecommand \bibitemNoStop [0]{.\EOS\space}%
\providecommand \EOS [0]{\spacefactor3000\relax}%
\providecommand \BibitemShut  [1]{\csname bibitem#1\endcsname}%
\let\auto@bib@innerbib\@empty
\bibitem [{\citenamefont {Wang}\ \emph {et~al.}(2020)\citenamefont {Wang},
  \citenamefont {Sciarrino}, \citenamefont {Laing},\ and\ \citenamefont
  {Thompson}}]{WangNatphoton2020}%
  \BibitemOpen
  \bibfield  {author} {\bibinfo {author} {\bibfnamefont {J.}~\bibnamefont
  {Wang}}, \bibinfo {author} {\bibfnamefont {F.}~\bibnamefont {Sciarrino}},
  \bibinfo {author} {\bibfnamefont {A.}~\bibnamefont {Laing}},\ and\ \bibinfo
  {author} {\bibfnamefont {M.~G.}\ \bibnamefont {Thompson}},\ }\bibfield
  {title} {\bibinfo {title} {Integrated photonic quantum technologies},\
  }\href@noop {} {\bibfield  {journal} {\bibinfo  {journal} {Nature Photonics}\
  }\textbf {\bibinfo {volume} {14}},\ \bibinfo {pages} {273–284} (\bibinfo
  {year} {2020})}\BibitemShut {NoStop}%
\bibitem [{\citenamefont {Moody}\ \emph {et~al.}(2022)\citenamefont {Moody}
  \emph {et~al.}}]{Moody_2022}%
  \BibitemOpen
  \bibfield  {author} {\bibinfo {author} {\bibfnamefont {G.}~\bibnamefont
  {Moody}} \emph {et~al.},\ }\bibfield  {title} {\bibinfo {title} {2022 roadmap
  on integrated quantum photonics},\ }\href
  {https://doi.org/10.1088/2515-7647/ac1ef4} {\bibfield  {journal} {\bibinfo
  {journal} {Journal of Physics: Photonics}\ }\textbf {\bibinfo {volume} {4}},\
  \bibinfo {pages} {012501} (\bibinfo {year} {2022})}\BibitemShut {NoStop}%
\bibitem [{\citenamefont {Rudolph}(2017)}]{RudlophOptimisticAPLP}%
  \BibitemOpen
  \bibfield  {author} {\bibinfo {author} {\bibfnamefont {T.}~\bibnamefont
  {Rudolph}},\ }\bibfield  {title} {\bibinfo {title} {Why {I} am optimistic
  about the silicon-photonic route to quantum computing},\ }\href@noop {}
  {\bibfield  {journal} {\bibinfo  {journal} {APL Photonics}\ }\textbf
  {\bibinfo {volume} {2}},\ \bibinfo {pages} {030901} (\bibinfo {year}
  {2017})}\BibitemShut {NoStop}%
\bibitem [{\citenamefont {Eli~Bourassa}\ \emph {et~al.}(2021)\citenamefont
  {Eli~Bourassa} \emph {et~al.}}]{XanaduBlueprint}%
  \BibitemOpen
  \bibfield  {author} {\bibinfo {author} {\bibfnamefont {J.}~\bibnamefont
  {Eli~Bourassa}} \emph {et~al.},\ }\bibfield  {title} {\bibinfo {title}
  {Blueprint for a scalable photonic fault-tolerant quantum computer},\
  }\href@noop {} {\bibfield  {journal} {\bibinfo  {journal} {Quantum}\ }\textbf
  {\bibinfo {volume} {5}},\ \bibinfo {pages} {392} (\bibinfo {year}
  {2021})}\BibitemShut {NoStop}%
\bibitem [{\citenamefont {Silverstone}\ \emph {et~al.}(2016)\citenamefont
  {Silverstone}, \citenamefont {Bonneau}, \citenamefont {O'Brien},\ and\
  \citenamefont {Thompson}}]{si-jstqe-22-390}%
  \BibitemOpen
  \bibfield  {author} {\bibinfo {author} {\bibfnamefont {J.~W.}\ \bibnamefont
  {Silverstone}}, \bibinfo {author} {\bibfnamefont {D.}~\bibnamefont
  {Bonneau}}, \bibinfo {author} {\bibfnamefont {J.~L.}\ \bibnamefont
  {O'Brien}},\ and\ \bibinfo {author} {\bibfnamefont {M.~G.}\ \bibnamefont
  {Thompson}},\ }\bibfield  {title} {\bibinfo {title} {Silicon quantum
  photonics},\ }\href@noop {} {\bibfield  {journal} {\bibinfo  {journal} {IEEE
  Journal of Selected Topics in Quantum Electronics}\ }\textbf {\bibinfo
  {volume} {22}},\ \bibinfo {pages} {390 } (\bibinfo {year}
  {2016})}\BibitemShut {NoStop}%
\bibitem [{\citenamefont {Giewont}\ \emph {et~al.}(2019)\citenamefont {Giewont}
  \emph {et~al.}}]{GiewontIEEEJSTQE2019}%
  \BibitemOpen
  \bibfield  {author} {\bibinfo {author} {\bibfnamefont {K.}~\bibnamefont
  {Giewont}} \emph {et~al.},\ }\bibfield  {title} {\bibinfo {title} {300-mm
  monolithic silicon photonics foundry technology},\ }\href@noop {} {\bibfield
  {journal} {\bibinfo  {journal} {IEEE J. Sel. Top. Quanutm Electron.}\
  }\textbf {\bibinfo {volume} {25}},\ \bibinfo {pages} {8200611} (\bibinfo
  {year} {2019})}\BibitemShut {NoStop}%
\bibitem [{\citenamefont {Silverstone}\ \emph {et~al.}(2014)\citenamefont
  {Silverstone} \emph {et~al.}}]{si-natphot-2014}%
  \BibitemOpen
  \bibfield  {author} {\bibinfo {author} {\bibfnamefont {J.~W.}\ \bibnamefont
  {Silverstone}} \emph {et~al.},\ }\bibfield  {title} {\bibinfo {title}
  {On-chip quantum interference between silicon photon-pair sources},\
  }\href@noop {} {\bibfield  {journal} {\bibinfo  {journal} {Nature Photonics}\
  }\textbf {\bibinfo {volume} {8}},\ \bibinfo {pages} {104 } (\bibinfo {year}
  {2014})}\BibitemShut {NoStop}%
\bibitem [{\citenamefont {Gentry}\ \emph {et~al.}(2015)\citenamefont {Gentry}
  \emph {et~al.}}]{GentryOptica2015}%
  \BibitemOpen
  \bibfield  {author} {\bibinfo {author} {\bibfnamefont {C.~M.}\ \bibnamefont
  {Gentry}} \emph {et~al.},\ }\bibfield  {title} {\bibinfo {title}
  {{Quantum-correlated photon pairs generated in a commercial 45~nm
  complementary metal-oxide semiconductor microelectronic chip}},\ }\href@noop
  {} {\bibfield  {journal} {\bibinfo  {journal} {Optica}\ }\textbf {\bibinfo
  {volume} {2}},\ \bibinfo {pages} {1065} (\bibinfo {year} {2015})}\BibitemShut
  {NoStop}%
\bibitem [{\citenamefont {Knoll}\ \emph {et~al.}(2015)\citenamefont {Knoll}
  \emph {et~al.}}]{ihp_epic}%
  \BibitemOpen
  \bibfield  {author} {\bibinfo {author} {\bibfnamefont {D.}~\bibnamefont
  {Knoll}} \emph {et~al.},\ }\bibfield  {title} {\bibinfo {title}
  {High-performance photonic bicmos process for the fabrication of
  high-bandwidth electronic-photonic integrated circuits},\ }in\ \href
  {https://doi.org/10.1109/IEDM.2015.7409706} {\emph {\bibinfo {booktitle}
  {2015 IEEE International Electron Devices Meeting (IEDM)}}}\ (\bibinfo {year}
  {2015})\ pp.\ \bibinfo {pages} {15.6.1--15.6.4}\BibitemShut {NoStop}%
\bibitem [{\citenamefont {Dziallas}\ \emph {et~al.}(2022)\citenamefont
  {Dziallas} \emph {et~al.}}]{epic1}%
  \BibitemOpen
  \bibfield  {author} {\bibinfo {author} {\bibfnamefont {G.}~\bibnamefont
  {Dziallas}} \emph {et~al.},\ }\bibfield  {title} {\bibinfo {title} {{A
  56-Gb/s Optical Receiver With 2.08-$\mu$A Noise Monolithically Integrated
  into a 250-nm SiGe BiCMOS Technology}},\ }\href
  {https://doi.org/10.1109/TMTT.2021.3104838} {\bibfield  {journal} {\bibinfo
  {journal} {IEEE Transactions on Microwave Theory and Techniques}\ }\textbf
  {\bibinfo {volume} {70}},\ \bibinfo {pages} {392} (\bibinfo {year}
  {2022})}\BibitemShut {NoStop}%
\bibitem [{\citenamefont {Gudyriev}\ \emph {et~al.}(2019)\citenamefont
  {Gudyriev} \emph {et~al.}}]{epic2}%
  \BibitemOpen
  \bibfield  {author} {\bibinfo {author} {\bibfnamefont {S.}~\bibnamefont
  {Gudyriev}} \emph {et~al.},\ }\bibfield  {title} {\bibinfo {title} {{Coherent
  ePIC Receiver for 64 GBaud QPSK in 0.25 $\mu$m Photonic BiCMOS Technology}},\
  }\href {https://doi.org/10.1109/JLT.2018.2881107} {\bibfield  {journal}
  {\bibinfo  {journal} {Journal of Lightwave Technology}\ }\textbf {\bibinfo
  {volume} {37}},\ \bibinfo {pages} {103} (\bibinfo {year} {2019})}\BibitemShut
  {NoStop}%
\bibitem [{\citenamefont {Rogers}\ \emph {et~al.}(2021)\citenamefont {Rogers}
  \emph {et~al.}}]{Rogers5123DimagerChip}%
  \BibitemOpen
  \bibfield  {author} {\bibinfo {author} {\bibfnamefont {C.}~\bibnamefont
  {Rogers}} \emph {et~al.},\ }\bibfield  {title} {\bibinfo {title} {A universal
  3d imaging sensor on a silicon photonics platform},\ }\href@noop {}
  {\bibfield  {journal} {\bibinfo  {journal} {Nature}\ }\textbf {\bibinfo
  {volume} {590}},\ \bibinfo {pages} {256} (\bibinfo {year}
  {2021})}\BibitemShut {NoStop}%
\bibitem [{\citenamefont {Lvovsky}\ \emph {et~al.}(2001)\citenamefont {Lvovsky}
  \emph {et~al.}}]{lvovsky2001quantum}%
  \BibitemOpen
  \bibfield  {author} {\bibinfo {author} {\bibfnamefont {A.~I.}\ \bibnamefont
  {Lvovsky}} \emph {et~al.},\ }\bibfield  {title} {\bibinfo {title} {Quantum
  state reconstruction of the single-photon fock state},\ }\href@noop {}
  {\bibfield  {journal} {\bibinfo  {journal} {Physical Review Letters}\
  }\textbf {\bibinfo {volume} {87}},\ \bibinfo {pages} {050402} (\bibinfo
  {year} {2001})}\BibitemShut {NoStop}%
\bibitem [{\citenamefont {Tse}\ \emph {et~al.}(2019)\citenamefont {Tse} \emph
  {et~al.}}]{PhysRevLett.123.231107}%
  \BibitemOpen
  \bibfield  {author} {\bibinfo {author} {\bibfnamefont {M.}~\bibnamefont
  {Tse}} \emph {et~al.},\ }\bibfield  {title} {\bibinfo {title}
  {Quantum-enhanced advanced ligo detectors in the era of gravitational-wave
  astronomy},\ }\href@noop {} {\bibfield  {journal} {\bibinfo  {journal}
  {Physical Review Letters}\ }\textbf {\bibinfo {volume} {123}},\ \bibinfo
  {pages} {231107} (\bibinfo {year} {2019})}\BibitemShut {NoStop}%
\bibitem [{\citenamefont {Acernese}\ \emph {et~al.}(2019)\citenamefont
  {Acernese} \emph {et~al.}}]{Acernese2019Dec}%
  \BibitemOpen
  \bibfield  {author} {\bibinfo {author} {\bibfnamefont {F.}~\bibnamefont
  {Acernese}} \emph {et~al.} (\bibinfo {collaboration} {Virgo Collaboration}),\
  }\bibfield  {title} {\bibinfo {title} {Increasing the astrophysical reach of
  the advanced virgo detector via the application of squeezed vacuum states of
  light},\ }\href {https://doi.org/10.1103/PhysRevLett.123.231108} {\bibfield
  {journal} {\bibinfo  {journal} {Phys. Rev. Lett.}\ }\textbf {\bibinfo
  {volume} {123}},\ \bibinfo {pages} {231108} (\bibinfo {year}
  {2019})}\BibitemShut {NoStop}%
\bibitem [{\citenamefont {Larsen}\ \emph {et~al.}(2019)\citenamefont {Larsen},
  \citenamefont {Guo}, \citenamefont {Breum}, \citenamefont
  {Neergaard-Nielsen},\ and\ \citenamefont {Andersen}}]{Larsen369}%
  \BibitemOpen
  \bibfield  {author} {\bibinfo {author} {\bibfnamefont {M.~V.}\ \bibnamefont
  {Larsen}}, \bibinfo {author} {\bibfnamefont {X.}~\bibnamefont {Guo}},
  \bibinfo {author} {\bibfnamefont {C.~R.}\ \bibnamefont {Breum}}, \bibinfo
  {author} {\bibfnamefont {J.~S.}\ \bibnamefont {Neergaard-Nielsen}},\ and\
  \bibinfo {author} {\bibfnamefont {U.~L.}\ \bibnamefont {Andersen}},\
  }\bibfield  {title} {\bibinfo {title} {Deterministic generation of a
  two-dimensional cluster state},\ }\href@noop {} {\bibfield  {journal}
  {\bibinfo  {journal} {Science}\ }\textbf {\bibinfo {volume} {366}},\ \bibinfo
  {pages} {369} (\bibinfo {year} {2019})}\BibitemShut {NoStop}%
\bibitem [{\citenamefont {Asavanant}\ \emph {et~al.}(2019)\citenamefont
  {Asavanant} \emph {et~al.}}]{Asavanant373}%
  \BibitemOpen
  \bibfield  {author} {\bibinfo {author} {\bibfnamefont {W.}~\bibnamefont
  {Asavanant}} \emph {et~al.},\ }\bibfield  {title} {\bibinfo {title}
  {Generation of time-domain-multiplexed two-dimensional cluster state},\
  }\href@noop {} {\bibfield  {journal} {\bibinfo  {journal} {Science}\ }\textbf
  {\bibinfo {volume} {366}},\ \bibinfo {pages} {373} (\bibinfo {year}
  {2019})}\BibitemShut {NoStop}%
\bibitem [{\citenamefont {Lodewyck}\ \emph {et~al.}(2007)\citenamefont
  {Lodewyck} \emph {et~al.}}]{LodewyckPRA2007CVQKD}%
  \BibitemOpen
  \bibfield  {author} {\bibinfo {author} {\bibfnamefont {J.}~\bibnamefont
  {Lodewyck}} \emph {et~al.},\ }\bibfield  {title} {\bibinfo {title} {{Quantum
  key distribution over 25 km with an all-fiber continuous-variable system}},\
  }\href@noop {} {\bibfield  {journal} {\bibinfo  {journal} {Phys. Rev. A}\
  }\textbf {\bibinfo {volume} {76}},\ \bibinfo {pages} {042305} (\bibinfo
  {year} {2007})}\BibitemShut {NoStop}%
\bibitem [{\citenamefont {Masada}\ \emph {et~al.}(2015)\citenamefont {Masada}
  \emph {et~al.}}]{ma-natphot-316-9}%
  \BibitemOpen
  \bibfield  {author} {\bibinfo {author} {\bibfnamefont {G.}~\bibnamefont
  {Masada}} \emph {et~al.},\ }\bibfield  {title} {\bibinfo {title}
  {Continuous-variable entanglement on a chip},\ }\href@noop {} {\bibfield
  {journal} {\bibinfo  {journal} {Nature Photonics}\ }\textbf {\bibinfo
  {volume} {9}},\ \bibinfo {pages} {316} (\bibinfo {year} {2015})}\BibitemShut
  {NoStop}%
\bibitem [{\citenamefont {Lenzini}\ \emph {et~al.}(2018)\citenamefont {Lenzini}
  \emph {et~al.}}]{Lenzinieaat9331}%
  \BibitemOpen
  \bibfield  {author} {\bibinfo {author} {\bibfnamefont {F.}~\bibnamefont
  {Lenzini}} \emph {et~al.},\ }\bibfield  {title} {\bibinfo {title} {Integrated
  photonic platform for quantum information with continuous variables},\
  }\href@noop {} {\bibfield  {journal} {\bibinfo  {journal} {Science Advances}\
  }\textbf {\bibinfo {volume} {4}} (\bibinfo {year} {2018})}\BibitemShut
  {NoStop}%
\bibitem [{\citenamefont {Raffaelli}\ \emph {et~al.}(2018)\citenamefont
  {Raffaelli} \emph {et~al.}}]{raffaelli2018homodyne}%
  \BibitemOpen
  \bibfield  {author} {\bibinfo {author} {\bibfnamefont {F.}~\bibnamefont
  {Raffaelli}} \emph {et~al.},\ }\bibfield  {title} {\bibinfo {title} {A
  homodyne detector integrated onto a photonic chip for measuring quantum
  states and generating random numbers},\ }\href@noop {} {\bibfield  {journal}
  {\bibinfo  {journal} {Quantum Science and Technology}\ }\textbf {\bibinfo
  {volume} {3}},\ \bibinfo {pages} {025003} (\bibinfo {year}
  {2018})}\BibitemShut {NoStop}%
\bibitem [{\citenamefont {Zhang}\ \emph {et~al.}(2019)\citenamefont {Zhang}
  \emph {et~al.}}]{zh-natphot-13-839}%
  \BibitemOpen
  \bibfield  {author} {\bibinfo {author} {\bibfnamefont {G.}~\bibnamefont
  {Zhang}} \emph {et~al.},\ }\bibfield  {title} {\bibinfo {title} {An
  integrated silicon photonic chip platform for continuous-variable quantum key
  distribution},\ }\href@noop {} {\bibfield  {journal} {\bibinfo  {journal}
  {Nature Photonics}\ }\textbf {\bibinfo {volume} {13}},\ \bibinfo {pages}
  {839} (\bibinfo {year} {2019})}\BibitemShut {NoStop}%
\bibitem [{\citenamefont {Tasker}\ \emph {et~al.}(2021)\citenamefont {Tasker},
  \citenamefont {Frazer} \emph {et~al.}}]{Tasker2021Jan}%
  \BibitemOpen
  \bibfield  {author} {\bibinfo {author} {\bibfnamefont {J.~F.}\ \bibnamefont
  {Tasker}}, \bibinfo {author} {\bibfnamefont {J.}~\bibnamefont {Frazer}},
  \emph {et~al.},\ }\bibfield  {title} {\bibinfo {title} {{Silicon photonics
  interfaced with integrated electronics for 9 GHz measurement of squeezed
  light - Nature Photonics}},\ }\href
  {https://doi.org/10.1038/s41566-020-00715-5} {\bibfield  {journal} {\bibinfo
  {journal} {Nat. Photonics}\ }\textbf {\bibinfo {volume} {15}},\ \bibinfo
  {pages} {11} (\bibinfo {year} {2021})}\BibitemShut {NoStop}%
\bibitem [{\citenamefont {Bruynsteen}\ \emph {et~al.}(2021)\citenamefont
  {Bruynsteen}, \citenamefont {Vanhoecke}, \citenamefont {Bauwelinck},\ and\
  \citenamefont {Yin}}]{Bruynsteen:21}%
  \BibitemOpen
  \bibfield  {author} {\bibinfo {author} {\bibfnamefont {C.}~\bibnamefont
  {Bruynsteen}}, \bibinfo {author} {\bibfnamefont {M.}~\bibnamefont
  {Vanhoecke}}, \bibinfo {author} {\bibfnamefont {J.}~\bibnamefont
  {Bauwelinck}},\ and\ \bibinfo {author} {\bibfnamefont {X.}~\bibnamefont
  {Yin}},\ }\bibfield  {title} {\bibinfo {title} {Integrated balanced homodyne
  photonic--electronic detector for beyond 20~ghz shot-noise-limited
  measurements},\ }\href {https://doi.org/10.1364/OPTICA.420973} {\bibfield
  {journal} {\bibinfo  {journal} {Optica}\ }\textbf {\bibinfo {volume} {8}},\
  \bibinfo {pages} {1146} (\bibinfo {year} {2021})}\BibitemShut {NoStop}%
\bibitem [{\citenamefont {Stojanovi\'{c}}\ \emph {et~al.}(2018)\citenamefont
  {Stojanovi\'{c}} \emph {et~al.}}]{soicmos}%
  \BibitemOpen
  \bibfield  {author} {\bibinfo {author} {\bibfnamefont {V.}~\bibnamefont
  {Stojanovi\'{c}}} \emph {et~al.},\ }\bibfield  {title} {\bibinfo {title}
  {{Monolithic silicon-photonic platforms in state-of-the-art CMOS SOI
  processes}},\ }\href {https://doi.org/10.1364/OE.26.013106} {\bibfield
  {journal} {\bibinfo  {journal} {Opt. Express}\ }\textbf {\bibinfo {volume}
  {26}},\ \bibinfo {pages} {13106} (\bibinfo {year} {2018})}\BibitemShut
  {NoStop}%
\bibitem [{\citenamefont {Feilchenfeld}\ \emph {et~al.}(2015)\citenamefont
  {Feilchenfeld} \emph {et~al.}}]{GFIBM}%
  \BibitemOpen
  \bibfield  {author} {\bibinfo {author} {\bibfnamefont {N.~B.}\ \bibnamefont
  {Feilchenfeld}} \emph {et~al.},\ }\bibfield  {title} {\bibinfo {title} {{An
  integrated silicon photonics technology for O-band datacom}},\ }in\ \href
  {https://doi.org/10.1109/IEDM.2015.7409768} {\emph {\bibinfo {booktitle}
  {2015 IEEE International Electron Devices Meeting (IEDM)}}}\ (\bibinfo {year}
  {2015})\ pp.\ \bibinfo {pages} {25.7.1--25.7.4}\BibitemShut {NoStop}%
\bibitem [{\citenamefont {Narasimha}\ \emph {et~al.}(2008)\citenamefont
  {Narasimha} \emph {et~al.}}]{luxtera}%
  \BibitemOpen
  \bibfield  {author} {\bibinfo {author} {\bibfnamefont {A.}~\bibnamefont
  {Narasimha}} \emph {et~al.},\ }\bibfield  {title} {\bibinfo {title} {{A
  40-Gb/s QSFP Optoelectronic Transceiver in a 0.13 $\mu$m CMOS
  Silicon-on-Insulator Technology}},\ }in\ \href
  {https://doi.org/10.1109/OFC.2008.4528356} {\emph {\bibinfo {booktitle}
  {OFC/NFOEC 2008 - 2008 Conference on Optical Fiber Communication/National
  Fiber Optic Engineers Conference}}}\ (\bibinfo {year} {2008})\ pp.\ \bibinfo
  {pages} {1--3}\BibitemShut {NoStop}%
\bibitem [{\citenamefont {Buckwalter}\ \emph {et~al.}(2012)\citenamefont
  {Buckwalter}, \citenamefont {Zheng}, \citenamefont {Li}, \citenamefont
  {Raj},\ and\ \citenamefont
  {Krishnamoorthy}}]{Monolithic15Oracle_Buckwalter_2012}%
  \BibitemOpen
  \bibfield  {author} {\bibinfo {author} {\bibfnamefont {J.~F.}\ \bibnamefont
  {Buckwalter}}, \bibinfo {author} {\bibfnamefont {X.}~\bibnamefont {Zheng}},
  \bibinfo {author} {\bibfnamefont {G.}~\bibnamefont {Li}}, \bibinfo {author}
  {\bibfnamefont {K.}~\bibnamefont {Raj}},\ and\ \bibinfo {author}
  {\bibfnamefont {A.~V.}\ \bibnamefont {Krishnamoorthy}},\ }\bibfield  {title}
  {\bibinfo {title} {{A Monolithic 25-Gb/s Transceiver With Photonic Ring
  Modulators and Ge Detectors in a 130-nm CMOS SOI Process}},\ }\href
  {https://doi.org/10.1109/JSSC.2012.2189835} {\bibfield  {journal} {\bibinfo
  {journal} {IEEE Journal of Solid-State Circuits}\ }\textbf {\bibinfo {volume}
  {47}},\ \bibinfo {pages} {1309} (\bibinfo {year} {2012})}\BibitemShut
  {NoStop}%
\bibitem [{\citenamefont {Cressler}(2013)}]{RetrospectiveOnSiGe_Cressler_2013}%
  \BibitemOpen
  \bibfield  {author} {\bibinfo {author} {\bibfnamefont {J.~D.}\ \bibnamefont
  {Cressler}},\ }\bibfield  {title} {\bibinfo {title} {{A retrospective on the
  SiGe HBT: What we do know, what we don't know, and what we would like to know
  better}},\ }in\ \href {https://doi.org/10.1109/SiRF.2013.6489439} {\emph
  {\bibinfo {booktitle} {2013 IEEE 13th Topical Meeting on Silicon Monolithic
  Integrated Circuits in RF Systems}}}\ (\bibinfo {year} {2013})\ pp.\ \bibinfo
  {pages} {81--83}\BibitemShut {NoStop}%
\bibitem [{\citenamefont {Knoll}\ \emph {et~al.}(2013)\citenamefont {Knoll}
  \emph {et~al.}}]{ihplocalsoi}%
  \BibitemOpen
  \bibfield  {author} {\bibinfo {author} {\bibfnamefont {D.}~\bibnamefont
  {Knoll}} \emph {et~al.},\ }\bibfield  {title} {\bibinfo {title} {{Substrate
  Design and Thermal Budget Tuning for Integration of Photonic Components in a
  High-Performance {SiGe}:C {BiCMOS} Process}},\ }\href
  {https://doi.org/10.1149/05009.0297ecst} {\bibfield  {journal} {\bibinfo
  {journal} {{ECS} Transactions}\ }\textbf {\bibinfo {volume} {50}},\ \bibinfo
  {pages} {297} (\bibinfo {year} {2013})}\BibitemShut {NoStop}%
\bibitem [{\citenamefont {Säckinger}(2010)}]{sackinger_limit}%
  \BibitemOpen
  \bibfield  {author} {\bibinfo {author} {\bibfnamefont {E.}~\bibnamefont
  {Säckinger}},\ }\bibfield  {title} {\bibinfo {title} {The transimpedance
  limit},\ }\href {https://doi.org/10.1109/TCSI.2009.2037847} {\bibfield
  {journal} {\bibinfo  {journal} {IEEE Transactions on Circuits and Systems I:
  Regular Papers}\ }\textbf {\bibinfo {volume} {57}},\ \bibinfo {pages} {1848}
  (\bibinfo {year} {2010})}\BibitemShut {NoStop}%
\bibitem [{\citenamefont {{Vokić}}\ \emph {et~al.}(2016)\citenamefont
  {{Vokić}}, \citenamefont {{Brandl}}, \citenamefont {{Schneider-Hornstein}},
  \citenamefont {{Goll}},\ and\ \citenamefont
  {{Zimmermann}}}]{VokicTIA_InputC}%
  \BibitemOpen
  \bibfield  {author} {\bibinfo {author} {\bibfnamefont {N.}~\bibnamefont
  {{Vokić}}}, \bibinfo {author} {\bibfnamefont {P.}~\bibnamefont {{Brandl}}},
  \bibinfo {author} {\bibfnamefont {K.}~\bibnamefont {{Schneider-Hornstein}}},
  \bibinfo {author} {\bibfnamefont {B.}~\bibnamefont {{Goll}}},\ and\ \bibinfo
  {author} {\bibfnamefont {H.}~\bibnamefont {{Zimmermann}}},\ }\bibfield
  {title} {\bibinfo {title} {10 gb/s switchable binary/pam-4 receiver and ring
  modulator driver for 3-d optoelectronic integration},\ }\href@noop {}
  {\bibfield  {journal} {\bibinfo  {journal} {IEEE Journal of Selected Topics
  in Quantum Electronics}\ }\textbf {\bibinfo {volume} {22}},\ \bibinfo {pages}
  {344} (\bibinfo {year} {2016})}\BibitemShut {NoStop}%
\bibitem [{\citenamefont {Lischke}\ \emph {et~al.}(2015)\citenamefont {Lischke}
  \emph {et~al.}}]{Lischke:15}%
  \BibitemOpen
  \bibfield  {author} {\bibinfo {author} {\bibfnamefont {S.}~\bibnamefont
  {Lischke}} \emph {et~al.},\ }\bibfield  {title} {\bibinfo {title} {High
  bandwidth, high responsivity waveguide-coupled germanium p-i-n photodiode},\
  }\href@noop {} {\bibfield  {journal} {\bibinfo  {journal} {Opt. Express}\
  }\textbf {\bibinfo {volume} {23}},\ \bibinfo {pages} {27213} (\bibinfo {year}
  {2015})}\BibitemShut {NoStop}%
\bibitem [{\citenamefont {Masalov}\ \emph {et~al.}(2017)\citenamefont
  {Masalov}, \citenamefont {Kuzhamuratov},\ and\ \citenamefont
  {Lvovsky}}]{masalov2017noise}%
  \BibitemOpen
  \bibfield  {author} {\bibinfo {author} {\bibfnamefont {A.~V.}\ \bibnamefont
  {Masalov}}, \bibinfo {author} {\bibfnamefont {A.}~\bibnamefont
  {Kuzhamuratov}},\ and\ \bibinfo {author} {\bibfnamefont {A.~I.}\ \bibnamefont
  {Lvovsky}},\ }\bibfield  {title} {\bibinfo {title} {Noise spectra in balanced
  optical detectors based on transimpedance amplifiers},\ }\href@noop {}
  {\bibfield  {journal} {\bibinfo  {journal} {Review of Scientific
  Instruments}\ }\textbf {\bibinfo {volume} {88}},\ \bibinfo {pages} {113109}
  (\bibinfo {year} {2017})}\BibitemShut {NoStop}%
\bibitem [{\citenamefont
  {S{\"{a}}ckinger}(2017)}]{Sackinger2017AnalysisReceivers}%
  \BibitemOpen
  \bibfield  {author} {\bibinfo {author} {\bibfnamefont {E.}~\bibnamefont
  {S{\"{a}}ckinger}},\ }\href {https://doi.org/10.1002/9781119264422} {\emph
  {\bibinfo {title} {Analysis and Design of Transimpedance Amplifiers for
  Optical Receivers}}}\ (\bibinfo  {publisher} {John Wiley {\&} Sons, Inc.},\
  \bibinfo {year} {2017})\ p.\ \bibinfo {pages} {208}\BibitemShut {NoStop}%
\bibitem [{\citenamefont {Liu}\ \emph {et~al.}(1999)\citenamefont {Liu},
  \citenamefont {Williams}, \citenamefont {Frankel},\ and\ \citenamefont
  {Esman}}]{Liu1999Jul}%
  \BibitemOpen
  \bibfield  {author} {\bibinfo {author} {\bibfnamefont {P.-L.}\ \bibnamefont
  {Liu}}, \bibinfo {author} {\bibfnamefont {K.~J.}\ \bibnamefont {Williams}},
  \bibinfo {author} {\bibfnamefont {M.~Y.}\ \bibnamefont {Frankel}},\ and\
  \bibinfo {author} {\bibfnamefont {R.~D.}\ \bibnamefont {Esman}},\ }\bibfield
  {title} {\bibinfo {title} {{Saturation characteristics of fast
  photodetectors}},\ }\href {https://doi.org/10.1109/22.775469} {\bibfield
  {journal} {\bibinfo  {journal} {IEEE Trans. Microwave Theory Tech.}\ }\textbf
  {\bibinfo {volume} {47}},\ \bibinfo {pages} {1297} (\bibinfo {year}
  {1999})}\BibitemShut {NoStop}%
\bibitem [{\citenamefont {Scott}\ and\ \citenamefont
  {Balram}(2022)}]{scott2022timing}%
  \BibitemOpen
  \bibfield  {author} {\bibinfo {author} {\bibfnamefont {J.~R.}\ \bibnamefont
  {Scott}}\ and\ \bibinfo {author} {\bibfnamefont {K.~C.}\ \bibnamefont
  {Balram}},\ }\bibfield  {title} {\bibinfo {title} {Timing constraints imposed
  by classical digital control systems on photonic implementations of
  measurement-based quantum computing},\ }\href@noop {} {\bibfield  {journal}
  {\bibinfo  {journal} {IEEE Transactions on Quantum Engineering}\ }\textbf
  {\bibinfo {volume} {3}},\ \bibinfo {pages} {1} (\bibinfo {year}
  {2022})}\BibitemShut {NoStop}%
\bibitem [{\citenamefont {Benedikovic}\ \emph {et~al.}(2019)\citenamefont
  {Benedikovic} \emph {et~al.}}]{Benedikovic:19}%
  \BibitemOpen
  \bibfield  {author} {\bibinfo {author} {\bibfnamefont {D.}~\bibnamefont
  {Benedikovic}} \emph {et~al.},\ }\bibfield  {title} {\bibinfo {title} {25
  gbps low-voltage hetero-structured silicon-germanium waveguide pin
  photodetectors for monolithic on-chip nanophotonic architectures},\
  }\href@noop {} {\bibfield  {journal} {\bibinfo  {journal} {Photon. Res.}\
  }\textbf {\bibinfo {volume} {7}},\ \bibinfo {pages} {437} (\bibinfo {year}
  {2019})}\BibitemShut {NoStop}%
\bibitem [{\citenamefont {Bakir}\ \emph {et~al.}(2010)\citenamefont {Bakir}
  \emph {et~al.}}]{BakirIEEE2010}%
  \BibitemOpen
  \bibfield  {author} {\bibinfo {author} {\bibfnamefont {B.}~\bibnamefont
  {Bakir}} \emph {et~al.},\ }\bibfield  {title} {\bibinfo {title} {Low-loss
  ($<$ 1 db) and polarization-insensitive edge fiber couplers fabricated on
  200-mm silicon-on-insulator wafers},\ }\href@noop {} {\bibfield  {journal}
  {\bibinfo  {journal} {IEEE Photonics Technology Letters}\ }\textbf {\bibinfo
  {volume} {22}},\ \bibinfo {pages} {739} (\bibinfo {year} {2010})}\BibitemShut
  {NoStop}%
\bibitem [{\citenamefont {Hamerly}\ \emph {et~al.}(2019)\citenamefont
  {Hamerly}, \citenamefont {Bernstein}, \citenamefont {Sludds}, \citenamefont
  {Solja{\v{c}}i{\'c}},\ and\ \citenamefont {Englund}}]{PhysRevX.9.021032}%
  \BibitemOpen
  \bibfield  {author} {\bibinfo {author} {\bibfnamefont {R.}~\bibnamefont
  {Hamerly}}, \bibinfo {author} {\bibfnamefont {L.}~\bibnamefont {Bernstein}},
  \bibinfo {author} {\bibfnamefont {A.}~\bibnamefont {Sludds}}, \bibinfo
  {author} {\bibfnamefont {M.}~\bibnamefont {Solja{\v{c}}i{\'c}}},\ and\
  \bibinfo {author} {\bibfnamefont {D.}~\bibnamefont {Englund}},\ }\bibfield
  {title} {\bibinfo {title} {Large-scale optical neural networks based on
  photoelectric multiplication},\ }\href@noop {} {\bibfield  {journal}
  {\bibinfo  {journal} {Physical Review X}\ }\textbf {\bibinfo {volume} {9}},\
  \bibinfo {pages} {021032} (\bibinfo {year} {2019})}\BibitemShut {NoStop}%
\bibitem [{\citenamefont {Vigliar}\ \emph {et~al.}(2021)\citenamefont {Vigliar}
  \emph {et~al.}}]{vi-natphys-17-1137}%
  \BibitemOpen
  \bibfield  {author} {\bibinfo {author} {\bibfnamefont {C.}~\bibnamefont
  {Vigliar}} \emph {et~al.},\ }\bibfield  {title} {\bibinfo {title}
  {Error-protected qubits in a silicon photonic chip},\ }\href@noop {}
  {\bibfield  {journal} {\bibinfo  {journal} {Nature Physics}\ }\textbf
  {\bibinfo {volume} {17}},\ \bibinfo {pages} {1137} (\bibinfo {year}
  {2021})}\BibitemShut {NoStop}%
\bibitem [{\citenamefont {Xiong}\ \emph {et~al.}(2016)\citenamefont {Xiong}
  \emph {et~al.}}]{Xiong2016Mar}%
  \BibitemOpen
  \bibfield  {author} {\bibinfo {author} {\bibfnamefont {C.}~\bibnamefont
  {Xiong}} \emph {et~al.},\ }\bibfield  {title} {\bibinfo {title} {{Active
  temporal multiplexing of indistinguishable heralded single photons}},\ }\href
  {https://doi.org/10.1038/ncomms10853} {\bibfield  {journal} {\bibinfo
  {journal} {Nat. Commun.}\ }\textbf {\bibinfo {volume} {7}},\ \bibinfo {pages}
  {1} (\bibinfo {year} {2016})}\BibitemShut {NoStop}%
\bibitem [{\citenamefont {Cable}\ and\ \citenamefont
  {Dowling}(2007)}]{CableDowlingPRLNOON2007}%
  \BibitemOpen
  \bibfield  {author} {\bibinfo {author} {\bibfnamefont {H.}~\bibnamefont
  {Cable}}\ and\ \bibinfo {author} {\bibfnamefont {J.~P.}\ \bibnamefont
  {Dowling}},\ }\bibfield  {title} {\bibinfo {title} {Efficient generation of
  large number-path entanglement using only linear optics and feed-forward},\
  }\href@noop {} {\bibfield  {journal} {\bibinfo  {journal} {Phys. Rev. Lett.}\
  }\textbf {\bibinfo {volume} {99}},\ \bibinfo {pages} {163604} (\bibinfo
  {year} {2007})}\BibitemShut {NoStop}%
\bibitem [{\citenamefont {Takeda}\ and\ \citenamefont
  {Furusawa}(2017)}]{Takeda2017Sep}%
  \BibitemOpen
  \bibfield  {author} {\bibinfo {author} {\bibfnamefont {S.}~\bibnamefont
  {Takeda}}\ and\ \bibinfo {author} {\bibfnamefont {A.}~\bibnamefont
  {Furusawa}},\ }\bibfield  {title} {\bibinfo {title} {{Universal Quantum
  Computing with Measurement-Induced Continuous-Variable Gate Sequence in a
  Loop-Based Architecture}},\ }\href
  {https://doi.org/10.1103/PhysRevLett.119.120504} {\bibfield  {journal}
  {\bibinfo  {journal} {Phys. Rev. Lett.}\ }\textbf {\bibinfo {volume} {119}},\
  \bibinfo {pages} {120504} (\bibinfo {year} {2017})}\BibitemShut {NoStop}%
\end{thebibliography}%

\noindent \textbf{Acknowledgements:} The authors thank K.~Balram, A.~Laing, J.~Silverstone and J.~Smith for insightful discussion. The authors are grateful for technical assistance from L.~Kling and A.~Murray and access to the University of Bristol Cleanroom. 
This work was supported by the Centre for Nanoscience and Quantum Information (NSQI), the European Research Council starting grant ERC-2018-STG 803665 and EPSRC Quantum Technology Capital fund: Quantum Photonic Integrated Circuits (QuPIC) EP/N015126/1. J.T. acknowledges support from an EPSRC Doctoral Training Partnership (1942717) for the experiment and device design stage of the project. J.C.F.M. acknowledges support from a Philip Leverhulme Prize.
\newline
\noindent \textbf{Data access statement:} The data and code that support the plots within this paper and other findings of this study are available from the corresponding author upon reasonable request.

\section{Appendix}

\subsection*{Current offset excess noise}
Due to the MMI imbalance before the photodiodes, we observe a power dependent net current offset, $i_{diff}$, at the amplifier input. This results in excess electronic noise at the amplifier output which we attribute to a combination of LO relative intensity noise and the amplifier's
DC current dependent gain. Figure A1 shows characterisation of the detector with symmetric 2~V reverse biases on each photodiode. 

\subsection*{Shot noise clearance limit}
The clearance of a balanced homodyne detector is described by a function of the form,
\begin{equation}
    \mathrm{SNC} = \frac{A}{B + Cf^2} + 1,
\end{equation}
where $A$ describes the shot noise contribution and $B$ and $C$ represent the white noise and quadratic noise terms of Equation 2 in the main text~\cite{masalov2017noise}. We fit this function to the measured clearance data, plotting the data and fit in Figure~A2.
\renewcommand{\thefigure}{A1}
\begin{figure}
    \centering
    \includegraphics{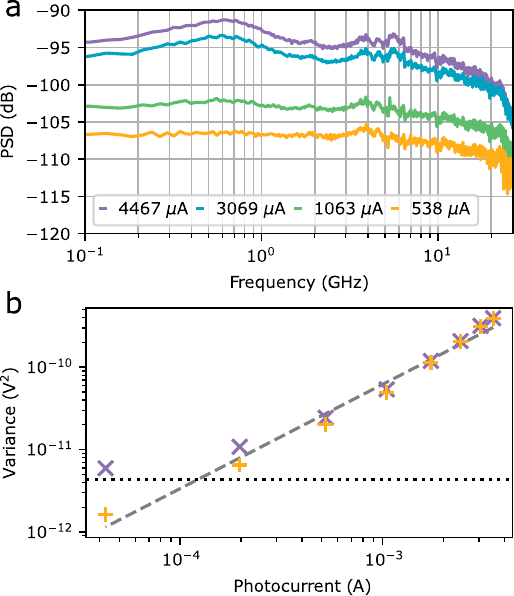}
    \caption{\textbf{Characterisation with imbalanced photocurrents.} \textbf{a}, PSD of the device response at different LO powers. ESA DANL, amplifier electronic noise and cable/ PCB transmission losses have been removed. We attribute the excess noise centred at 6~GHz to intensity noise from the EDFA. \textbf{b}, Raw (purple crosses) and electronic noise subtracted (orange pluses) noise variances against total photocurrent. Dashed lines indicate the electronic noise level and a linear fit to the data, respectively. The fit gives a gradient of $1.26\pm0.01$, indicating the presence of excess classical noise in addition to vacuum shot noise.} 
    \label{fig:hd5_suppl_characterisation}
\end{figure}
\renewcommand{\thefigure}{A2}
\begin{figure}1
    \centering \includegraphics{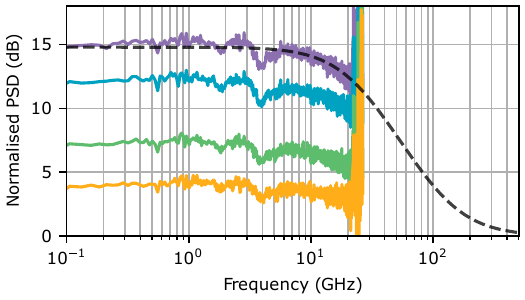}
    \caption{\textbf{Shot noise clearance fit.} We normalise the measured detector shot-noise response to the amplifier and spectrum analyser electronic noise to obtain the ratio of quantum to classical noise, or clearance. We extrapolate the trend beyond our 26.5~GHz measurement bandwidth through a fit to Equation 1, suggesting clearance beyond 100~GHz.} 
    \label{fig:hd5_clearance_extrap}
\end{figure}


\end{document}